# A Review of Recent Developments in Atomic Processes for Divertors and Edge Plasmas

D. E. Post

ITER, San Diego Co-Center, San Diego, CA, USA

**Abstract**

The most promising concepts for power and particle control in tokamaks and other fusion experiments rely upon atomic processes to transfer the power and momentum from the edge plasma to the plasma chamber walls. This places a new emphasis on processes at low temperatures (1-200 eV) and high densities ($10^{20}$—$10^{22}$ m$^{-3}$). The most important atomic processes are impurity and hydrogen radiation, ionization, excitation, recombination, charge exchange, radiation transport, molecular collisions, and elastic scattering of atoms, molecules and ions. Important new developments have occurred in each of these areas. New calculations for impurity and hydrogen ionization, recombination, and excitation rate coefficients for low temperature plasmas that include collisional radiative effects and multi-configuration interactions indicate that earlier estimates of these rate coefficients for plasmas with low Z impurities such as beryllium are too high in selected regions of interest[1]. Transport effects and charge exchange recombination are also key elements for determining the ionization-recombination balance and radiation losses. Collisional radiative effects for hydrogen are an essential ingredient for understanding and modelling high recycling divertors[2, 3]. Since the opacity in a high recycling divertor can be large for many of the strongest lines of recycling hydrogen atoms and low Z impurities, the transport of this radiation determines where the energy is deposited and strongly influences the recombination and ionization rate coefficients[4, 5]. Molecular collisions play a large role in recycling and in the energy and particle balance at the plasma edge[6]. Charge Exchange and elastic collisions are also important for determining the neutral gas pressure and the transfer of plasma energy and momentum to the chamber walls[7, 8]. The best available data for these processes and an assessment of their role in plasma wall interactions are summarized, and the major areas where improved data are needed are reviewed.

## 1. Introduction

Atomic and molecular processes play a key role in divertors. The particle balance in divertors is determined by the recycling of plasma and neutral atoms and molecules. The plasma momentum balance is influenced by "friction" with the neutrals due to charge exchange and elastic collisions. Impurity radiation, hydrogen radiation, ionization, charge exchange collisions, elastic scattering, and radiation transport can all play key roles in the energy balance in the divertor.

The alpha particle heating power for ITER is expected to be 300 to 600 MW for pulse lengths as long as 1000 s. The peak heat loads will be very high, in the range of 20 to 40 MW/m$^2$ or more[9, 10]. Directly terminating the plasma on metal divertor plates will not be adequate because the effective "wetted area" of the divertor plates will be too small to lead to acceptably low heat fluxes. The peak heat loads can be reduced to the 0.6—4 MW/m$^2$ range if the heat load can be spread out on the walls of the divertor chamber. The present concept being developed for the ITER divertor is designed to maximize the role of atomic processes such as charge exchange, hydrogen and impurity line radiation, ionization, and elastic collisions between the recycling gas and the plasma in the diverted plasma to spread out the heat and momentum. Conditions where atomic processes have dispersed the heat and momentum have been realized on a number of tokamaks, including PDX, DIII-D, JT-60U, ASDEX/U and JET[12, 13, 14, 15], but with lower power levels than needed for a next step experiment such as ITER. Present experiments will not be able to test the physics of divertor operation for these high powers, so that such a divertor concept must be based on extrapolations of data from present experiments using models validated on the present experiments. These models must include as much of the important physical effects as possible and will therefore require an accurate treatment of the important atomic processes.

Present divertor experiments such as DIII-D can be used to define the appropriate conditions for atomic processes. The conditions for a detached plasmas in DIII[11] and Alcator C-mod[12] indicate that the density at the divertor plate and the main plasma edge can be very similar ( $\sim 10^{20}$ m$^{-3}$), but that the temperatures drop from $\sim$ 100—200 eV at the boundary of the main plasma to 5—10 eV or lower at the divertor plate.

The atomic physics important for divertor plasmas can be divided into the physics of hydrogen ions, atoms and molecules, and the physics of heavier atoms and ions (impurities). Hydrogen recycling plays an important role in the particle and pressure/momentum balance, and can play a large role in the energy balance. However it is very difficult to exhaust most of the energy from the divertor plasma using hydrogen processes. The conditions suitable for the transfer of the energy from the plasma to the divertor sidewalls with impurity radiation are less severe than for hydrogen radiation and fast particles, so that impurity radiation offers the more promise for energy exhaust than do hydrogen based processes. All of these considerations need to be assessed using very accurate atomic data. Over the last 15 years, the progress in producing and collecting such data has been substantial, both for hydrogen atoms and molecules and impurities and will be reviewed in the paper.

## II. HYDROGEN ATOMS AND MOLECULES

The recycling of hydrogen atoms and molecules plays a crucial role in the particle and momentum transport and can play a very significant role in energy transport as well. The flux of ions along the field lines is determined by ionization and recombination in the continuity equation (Eq. 1).

$$\frac{\partial(nv_\parallel)}{\partial x} = <\sigma v>_{ionization} n_e n_o - <\sigma v>_{recombination} n_e n_i \qquad (1)$$

For the plasmas being considered with $n_e \geq 10^{19}$ m$^{-3}$, the time between electron excitation, de-excitation and ionization collisions is comparable to the radiative decay time for the excited states of hydrogen so that "multi-step" effects are important. The rate coefficients therefore depend on density. For $T_e \leq 3$ eV, three body recombination is also important. For these reasons, a "collisional radiative" treatment is necessary for hydrogen[2, 3, 13]. These effects increase the ionization and recombination rate coefficients and decrease the radiation rate coefficients at high densities. As noted earlier, reducing the plasma temperature to a few eV is not sufficient to reduce the heat flux because the plasma deposits the ionization energy (13.6 eV per ion-electron pair) on the plate as it recombines. It is essential that the energy be transferred to the plasma facing components through photons and low energy particles. Volume recombination of the divertor plasma before it reaches the divertor plate would reduce the particle and energy fluxes to the divertor plate. The ionization and recombination rate coefficients are approximately equal at about 1.3 eV (Figure 1). Collisional radiative effects strongly depress the hydrogen radiation as well (Figure 2). Even with this suppression, a number of calculations using 2-D plasma simulations[14] have found that, at very low temperatures with a strongly recombining plasma, hydrogen radiation can radiate all of the power to the walls. For these cases the electron temperature is very low in the recombining region ($\leq 1$ eV) and the density of neutral atoms is very high ($\geq 10^{22}$ m$^{-3}$). The recombination rate is relatively low ($\leq 10^{-11}$ cm$^3$ s$^{-1}$) so that for $n_e \sim 10^{14}$ cm$^{-3}$, $\tau_{recomb} = 10^{-3}$ s. This is comparable to or longer than the ion recycle time in the divertor for a flow speed of $\sim 10^6$ cm/s Therefore, the plasma density must be very high and the flow speed very small for three-body and radiative recombination to have an important effect.

For $n_H \geq 10^{19}$ m$^{-3}$, the neutral cloud is opaque to much of the hydrogen radiation (Table 1) (e.g. Krashenninikov, et al[5, 15]). For Doppler broadened lines:

$$\lambda_{absorption} = b_{ul} \left( \frac{T_i(eV)}{\mu_i} \right)^{.5} \frac{1}{n_{20}} meters \qquad (2)$$

Table 1 Parameters for a number of the H transitions of interest with $n_o \sim 10^{20}$ m$^{-3}$.[16]

| Transition | $A_{ul}$ | λ (nm) | $b_{ul}$ | $\lambda_{absorption}$ |
|---|---|---|---|---|
| 2 —> 1 | $4.7 \times 10^8$ | 127 | $1.6 \times 10^{-3}$ | 0.2 cm |
| 3 —> 2 | $4.4 \times 10^7$ | 656 | $2.2 \times 10^{-4}$ | 0.04 cm |
| 3 —> 1 | $5.5 \times 10^7$ | 102 | $1.2 \times 10^{-2}$ | 2 cm |

For typical transitions such as $L_\alpha$, $L_\beta$, $H_\alpha$, $n_0 \sim 10^{20\text{—}22}$ m$^{-3}$ and $T_i \sim 3$ eV, $\lambda_{absorption}$ can be small compared to the dimensions of the neutral cloud. Assessment of the effect on the ionization, recombination, and hydrogen radiation loss rate coefficients requires a treatment of the transport of the line radiation. $\lambda_{absorption}$ is small at the wavelength of the lines and large at other wavelengths. The lines are Doppler shifted and broadened and pressure broadened as well. As outlined in [3], an estimate can be made of the magnitude of the effect by suppressing the Lyman transitions (n > 1 ↔ n = 1). For this case, the recombination and ionization rate coefficients cross at a slightly lower temperature (0.9 eV compared to 1.2 eV), and the ionization rate coefficients are roughly equal to those calculated for high densities with all transitions included (Figure 3). Thus the plasma recombination/ionization balance will change if radiation trapping is included, but probably not drastically. Detailed calculations including these effects will be necessary to determine the ionization-recombination balance. These effects would influence where the hydrogen radiation is deposited on the plasma facing components. Much of the radiation would be reflected back toward the main plasma. The net radiation rate is also lowered(Figure 4). Detailed calculations will be necessary to determine the detailed ionization/recombination balance.

    A. Wan et al [4] have used a non-LTE collisional radiative model (GLF and CRETIN[17])which includes two dimensional transport of the hydrogen line and continuum radiation to make a preliminary assessment of the role of hydrogen line transport in a model plasma based on the DIII-D detached plasma discussed in an earlier session[18]. The model included the energy levels up to n=10. They compared three cases, (1) Lyman α radiation assuming that the plasma was thin to the neutrals, and (2) full treatment of the Lyman α transport, and (3) detailed line transfer for all lines up to n=6. The effect on the relative ion and neutral populations is illustrated in Figure 5 for the first two cases. The ionization/recombination balance is affected by the absorption and re-emission of the Lyman α lines. The results for the hydrogen line energy fluxes are summarized in Table 2. For the densities observed in the DIII-D detached plasmas, inclusion of the neutral opacity produces a significant change in the distribution of the radiation.

  Table 2 Ratio of Hydrogen radiation fluxes to the divertor plate and back to the main plasma[4]

|                                              | back to core | to divertor plate |
|----------------------------------------------|--------------|-------------------|
| Lyman-α without line opacity                 | 0.428        | 0.427             |
| detailed line transfer of Lyman-α            | 0.146        | 0.071             |
| detailed line transfer of many hydrogen lines| 0.164        | 0.091             |

At temperatures of 1—10 eV in the divertor plasma, molecular effects will be important. $H_2$ will be formed at the divertor walls from the atomic hydrogen and ions incident on the plasma facing components. Much of the $H_2$ will be vibrationally excited due to desorption from the wall, collisions with ions, neutrals and electrons, and charge exchange with $H_2^+$[19]. The dissociation and ionization rate coefficients of vibrationally excited $H_2$ are much larger than ground state $H_2$ molecules. Many new reaction channels open up. For instance it has been suggested as early as 1984 [2, 20] that, at low temperatures, a large density of vibrationally excited $H_2$ could enhance the recombination rate of $H^+$ through the reaction chain:

$$H_2(v=0) + p, e^-, wall, H_2^+ \Rightarrow H_2(v \geq 0)$$
$$H_2(v \geq 0) + e^- \Rightarrow H^- + H^o \qquad (3)$$
$$H^- + H^+ \Rightarrow H^o + H^o(n=3)$$

The charge exchange recombination rate (Eq. 3) is about $4 \times 10^{-8}$ cm$^3$ s$^{-1}$ (Figure 6), about four orders of magnitude larger than the radiative and three body recombination rate coefficients at these temperatures. The recombination time would thus be about $1 / (10^{14}$ cm$^{-3} \times 4 \times 10^{-8}$ cm$^3$ s$^{-1}) \approx$ 2.5×10$^{-7}$ s, compared to 5×10$^{-3}$ s for radiative and three body recombination. Since transit times in the divertor are of the order of several × 10$^{-3}$ s for ions, unless the flow is very stagnated and the density is very high, there is usually not sufficient time for volume recombination to occur. The formation of H⁻ competes with dissociation, but is dominant for temperatures ≤ 5 eV. Because the binding energy of H⁻ is 0.75 eV, destruction of H⁻ is a rapid process for $T_e$ above ~ 1 eV so that this recombination mechanism is most effective near 1 eV and below. These effects have played a large role in negative ion sources[21, 22]. Measurements of negative ion fractions as high as 50% for neutral pressures of 2 mTorr have been observed[23] in cusp discharges used in negative ion source research(Figure 7). This mechanism could be important in detached plasmas because it would allow the plasma to partially recombine before it reaches the divertor plate which could contribute to the decrease in the plasma flux on the divertor plate observed in many detached plasma experiments[11, 12]. $H_2$ formed on the walls can also be emitted in a vibrationally excited state which would speed the formation process.

Hydrogen processes are also potentially important for momentum balance. Computational models[24] and analytic models[8, 25] indicate that charge exchange friction can be an important contributor to the momentum balance.

$$\frac{\partial}{\partial x}\left(nmv_x^2 + p\right) = -R_{cx+recomb} \quad \vdots \quad R_{cx+recomb} \equiv \left(\langle\sigma v\rangle_{cx} n_o + \langle\sigma v\rangle_{recomb} n_e\right) n_i \, m |v_x| \qquad (4)$$

For charge exchange and other ion neutral collision processes to be important, it is necessary that the neutrals be able to penetrate the plasma, undergo a collision, and leave the plasma transferring the momentum (and some energy) to the walls many times before being ionized. This requires that the electron temperature be in the 2—4 eV range so that $\lambda_{ioniz} \gg \lambda_{CX}$. Including collisional radiative enhancements [2] to the ionization rate coefficients decreases the temperature at which the ionization rate coefficient is 10% or less of the total reaction rate(Figure 8).

Elastic scattering of H and $H_2$ with each other and with the plasma ions is also potentially very important. The cross sections are large[26, 27, 28], primarily in the forward direction so that calculations of the transport, momentum and energy transfer rates for collisions such as $H + H$, $H + H_2$, $H + H^+$ and $H_2 + H^+$ are needed. Calculations using classical cross sections are being done with Monte Carlo neutral transport codes[27, 28, 29]. Elastic collisions have the potential for reflecting $H_2$ from the plasma [28, 29], retarding the penetration of neutrals into the plasma, and providing a buffer and a path for heat conduction between the plasma and the divertor chamber walls. They can also contribute to the momentum loss rate. To the extent that they affect the neutral pressure, they can affect the pumping rate. The effective elastic scattering cross section for atomic hydrogen is $\sim 3 \times 10^{-19}$ m$^2$, so that for a neutral density of $10^{20}$ m$^{-3}$, the mean free path is about 0.03 m, less than the distance to the walls of the divertor chamber.

## 3. Impurity Radiation:

Impurity radiation appears to offer the most promise for exhausting the power to the divertor chamber walls. Impurity radiation plays a major, and probably the dominant role in the energy losses in detached plasmas[12, 18]. Although the measurement of impurity radiation and the identification of the impurity species is crucial for understanding impurity behavior in the divertor, the present discussion of impurity radiation is restricted to the calculation of the total radiated power. The status and issues for spectroscopic measurements of impurities in the plasma edge has recently been reviewed by Wiese[30, 31], Bitter[32], and von Hellerman[33]. Radiation losses due to line emission, especially from metals, were recognized to be very important energy loss channels in the central plasma of tokamaks in the 1970's [34]. The early calculations for impurity radiation rate coefficients used very simple models based on semi-analytic hydrogenic prescriptions for the energy levels and rate coefficients for an "average ion" configuration[35] and were designed for highly charged high Z ions at temperatures of a few keV or higher. Partially due to the interest and support of the fusion community as well as the astrophysical community, considerable progress has been made over the last 20 years in the understanding of atomic processes important in fusion plasmas. Much of this focus has been on conditions appropriate for core plasmas, temperatures of 1 to 10 keV, and densities of $10^{19}$—$10^{20}$ m$^{-3}$. The divertor plasmas will be much colder and potentially have regions as dense as $10^{21}$ and perhaps $10^{22}$ m$^{-3}$. The original models for the atomic cross sections and rate coefficients were based on relatively simple scalings from a few experiments and a few calculations that were very limited compared to present experiments and to the calculations that can be performed with today's computers[36, 37, 38, 39]. Since then, there have been extensive measurements of the basic processes and even more extensive calculations of the collision cross sections and other processes (c.f. [40]). Because there are a large number of ions, each with many levels and configurations, it is not practical to measure all of the cross sections and rate coefficients. The approach has been to use measurements of key cross sections and rate coefficients to identify the major physics elements and to provide quantitative benchmarks for theoretical calculations[41]. This has been a very successful strategy, resulting in agreement between theory and experiment often better than 10 to 20%.

Impurity radiation is due to radiative decay of excited states of the various charge states of the impurity ions. The distribution of charge states is determined by transport, ionization, and recombination.

$$\frac{\partial n_z^{+i}}{\partial t} + \nabla \bullet \Gamma_z^{+i} =$$
$$n_e n_z^{+i-1} <\sigma v>_{ioniz}^{+i-1 \to +i} - n_e n_z^{+i}(<\sigma v>_{ioniz}^{+i \to i+1} + <\sigma v>_{recomb}^{+i \to i-1}) + n_e n_z^{+i+1} <\sigma v>_{recomb}^{+i+1 \to i} \quad (5).$$
$$P_{radiation} = \sum_{i=0}^{z} n_e n_z^{+i} \sum^{all\_l} <\sigma v>_{exicitation}^{g \to l} (E_l - E_g)$$

Early estimates of ionization cross sections were largely based on semi-empirical scalings [42] and there were relatively few measurements[43] to benchmark the scalings. The development of plasma devices which can produce beams of multiply charged ions using ECRH, high energy (multi-keV) electron beams, and other techniques have resulted in extensive measurements of the direct ionization cross sections for most of the lighter ions[41]. These measurement are used to guide and benchmark multi-configuration calculations of ionization which usually agree with the experiments to a high degree of accuracy. The calculation of accurate cross sections for direct ionization has been more difficult than is first apparent. Although usually only a few configurations are needed and there are only one initial and one final state for the ion and the wave functions of the ingoing and outgoing electrons are simple, there are two outgoing electrons in addition to the impurity ion, making the final state a three body system. Indirect processes, involving the production of intermediate excited states, also contribute to the total ionization cross section. In spite of these complications, the degree of agreement between theory and experiment today is very good[44, 45], often to within 10 or 20%. Figure 9 compares the measured and calculated cross sections for the indirect and direct ionization of Scandium III($Sc^{2+}$) to Scandium $Sc^{3+}$)[46] and the Lotz semi-empirical scaling for direct ionization[48]. The contribution to the cross section from the resonant excitation of an inner shell electron ($3p^6$) to higher levels leading to Auger ionization is a factor of 10 higher than the direct ionization. The ionization cross sections for low Z ions have generally been measured and calculated, but measurements are needed for the low ionization states of metals, particularly heavy metals such as Mo and W [41].

Three types of recombination are important in tokamaks: radiative, di-electronic, and charge transfer with hydrogen neutrals. Radiative recombination is a relatively simple process which can be calculated with a high degree of accuracy. Even for this relatively simple process, however, refinements are still being made. It is, however, usually the weakest of the three. Di-electronic recombination is normally much stronger[47]. Di-electronic recombination involves the simultaneous excitation of a bound electron and the capture of the incident electron in a bound state in the doubly excited ion. One of the excited electrons can then radiatively decay to a lower energy level, resulting in a singly excited ion which eventually decays to the ground state. Di-electronic recombination cross section measurements are much more difficult than ionization measurements because they involve the sum of many resonant processes, each of which makes a small contribution to the total rate. The cross sections must be measured using merged beams to obtain small relative velocities between the reactants and, in addition, the low relative velocity leads to low reaction rate coefficients. In addition charge transfer between the background gas atoms and the multi-charged ion can be a much stronger process than the specific recombination event being measured. The theory is also difficult because many states are needed for an accurate calculation of the excitation rate coefficients and accurate energy levels are required. However, measurements of di-electronic recombination cross sections are now being made on large storage rings which can store beams of multiply charge ions (e.g. $C^{6+}$, $O^{6+}$, $Ar^{10+}$, $Cu^{26+}$, etc.). The calculations are in good agreement with the experiments (Figure 10).

The third type of recombination is due to charge exchange between hydrogen atoms and impurity ions. This process has been observed to be important for radiation losses in tokamaks which have been heated with neutral beams[48], and is extensively exploited to measure the ion temperature, plasma rotation velocities, fully charged impurity densities, and density fluctuations in tokamaks [33, 49, 50, 51]. Early measurements and theories for the cross sections were done for relative energies typical of neutral beams used for plasma heating ( 20—200 keV/amu) (c.f. [52]). A common scaling for the cross sections is fairly simple ($\sigma \approx 10^{-15}$ cm$^2$/q, where q is the ionic charge)[53]. Charge exchange recombination can be important for enhancing the radiation losses in

Shortened PSI Review Paper 6    10

edge plasmas[54] and considerable effort has devoted to the measurement and calculation of the cross sections for the low relative energies characteristic of edge plasmas (1—100 eV/amu). These measurements are more difficult than the high energy measurements because high current, low energy beams of multiply charged ions are more difficult to form than high energy beams. The calculations are more difficult as well because, at low relative energies, the collision is not a simple binary collision with a rapid interaction between the two colliding particles. The impurity ion and hydrogen atom are close to each other long enough for a "quasi-molecule" to form which means that the energy levels and configurations of the "molecular" state much be calculated. The new generation of multi-charged ion sources have allowed the measurement of cross sections down to relative energies of 10 eV/amu. Detailed calculations including all of the molecular effects are in good agreement with the experiments. An assessment of the available data most relevant to the plasma edge has been recently made by Phaneuf[53].

Because impurity radiation losses are almost entirely due to the radiative decay of excited states produced by electron impact excitation of impurity ions, it is very important to be able to calculate accurate excitation rate coefficients. Initially, excitation cross sections were measured by crossed beams of a multi-charged ions and electrons and measuring the photons from the radiative decay of the excited state that is produced[36, 55]. This is an extremely difficult measurement because the photons are radiated istropically into $4\pi$ steradians, and the cross section for a particular excitation is often small (~ $10^{-22}$ m$^2$) [43]. The cross sections are now measured by measuring the energy loss of electrons in a probe beam of mono-energetic electrons crossing a target beam of multi-charged ions[56]. The calculations, which are even more important because it is not practical to measure all of the millions of possible excitations, have also become very accurate (c.f. [57]) and the agreement between experiments and theory is generally excellent[56, 58] (Figure 11).

Many of the original calculations for impurity ionization, recombination and radiation rate coefficients used semi-analytic fits for the rate coefficients and screened hydrogenic models for the electronic configurations and energy levels. These were embodied in the ADPAK set of rate coefficients [35, 59] which is still widely used because the package is convenient, is reasonably accurate, and has rate coefficients for all charge states and elements from hydrogen to uranium (and beyond). For a particular species, however, for which data for a specific charge state of a particular is available, this package does not supply the best available rates. Its main virtue is that it is accessible to the people who need a rate package. The main elements of this package date back almost 20 years and much better calculations for many low Z elements are available. Improved descriptions of the energy levels, meta-stable levels, much better ionization and recombination rate coefficients, and much better excitation rate coefficients have been included in a full collisional radiative model[1, 60]. Figure 12 shows the difference that various models[1] make in the emission rate coefficients and compares them to the ADPAK [59]rate coefficients. The distorted wave calculations use better wave functions than the plane wave Born calculations, and the detailed level calculation has a better treatment of the energy levels and electron configurations than the configuration average calculation. The most accurate calculations(Detailed Levels, Distorted Wave) for the emission rates below 10 eV are a factor of 5 to 10 lower than the simplest model (Configuration Average, Plane Wave Born). The collisional radiative treatment is also an important feature of the calculations. Due to the presence of meta-stables and other long lived states, electron collisions can de-excite or ionize excited electrons before they decay radiatively. In addition, multiple electron collisions can reduce the di-electronic recombination rate coefficients. These effects are strongest for low Z atoms and ions (Figure 13)[1].

    The major challenge confronting the use of impurity radiation to transfer the power from the divertor plasma to the divertor side-walls is to obtain a sufficiently high density of the right impurity in the divertor chamber to radiate the power while avoiding excessive contamination of the main plasma. The allowed impurity concentration depends very strongly on the atomic number. Figure 14 shows the impurity radiation rate coefficients for a number of candidate impurities being considered for use in divertors. The appropriate choice of impurity species depends on the desired temperature range. Increased Z, increased plasma density, and increased impurity concentration all increase the power loss due to impurity radiation. The impurity content of the main plasma must be

less than the "fatal fraction" above which ignition becomes impossible due to dilution and radiation losses from the central plasma(Table 3)[61]. If there is a net hydrogen flow toward the divertor chamber, friction forces with the background plasma would tend to retain the impurities in the divertor[62]. However, the thermal force tends to lead to impurity transport toward the main plasma, and drifts, two dimensional and three dimensional flows in the scrape-off layer, especially near the X point, and other effects vastly complicate the situation, so it is prudent to assume that the impurity concentration in the divertor is no higher than the central concentration.

Table 3  Maximum Allowed Impurity Fractions for Ignition[61]

| Z | 4 | 6 | 10 | 18 | 26 | 42 | 74 |
|---|---|---|----|----|----|----|----|
| "Fatal Fraction" | 14% | 6.7% | 2.4% | 0.54% | 0.25% | 0.11% | 0.011% |

The heating power must be transported from the main plasma to the radiating region in the divertor. Impurity radiation is most effective when there are bound electrons in partially filled shells. Fully stripped, He-like, Ne-like, etc. ions do not radiate strongly[35]. This means that the temperature range over which the impurities radiate can be very limited, especially for low Z impurities. Since heat conduction requires a temperature gradient, the range in $T_e$ for a given impurity may not be large enough to obtain adequately large radiation losses. As pointed out by Rebut and Green[63], Lengyel[64] and K. Lackner[64, 65], the total radiation losses depend only on temperature if the major portion of the heat transport is due to electron conduction. This can be derived assuming pressure balance along the field lines (equation 6).

$$\frac{\partial Q_\parallel}{\partial x} = -n_e n_z L_Z(T_e) \;\vdots\; Q_\parallel = -\kappa_o T_e^{2.5} \frac{\partial T_e}{\partial x} \;\vdots\; p_e = n_e T_e \Rightarrow$$

$$Q_\parallel \frac{\partial Q_\parallel}{\partial x} = -n_e n_z L_Z Q_\parallel \approx \kappa_o n_e n_z L_Z T_e^{2.5} \frac{\partial T_e}{\partial x} \Rightarrow$$

$$\frac{1}{2}\frac{\partial Q_\parallel^2}{\partial x} \approx \frac{p_e^2}{T_e^2}\kappa_o f_z L_Z T_e^{2.5}\frac{\partial T_e}{\partial x} \approx p_e^2 \kappa_o f_z L_Z T_e^{0.5}\frac{\partial T_e}{\partial x} \Rightarrow$$

$$\frac{1}{2}dQ_\parallel^2 \approx p_{es}^2 \kappa_o f_z L_Z T_e^{0.5} dT_e \Rightarrow \frac{\Delta Q_\parallel}{n_{es}\sqrt{F_z}} \approx \sqrt{2\overline{\kappa}_o T_{es}^2 \int_0^{T_{es}} L_Z(T_e) T_e^{0.5} dT_e}$$

where $\kappa_o \approx \dfrac{3.1\times 10^9}{Z_{eff} \ln \Lambda}\left(\dfrac{erg}{cm\, s\, eV^{3.5}}\right)$ and $F_z(f_Z) \equiv \dfrac{f_Z(\%)}{Z_{eff}} = \dfrac{f_Z(\%)}{1+.01 f_Z(\%) Z(Z-1)}$

with $\overline{\kappa}_o \equiv \kappa_o Z_{eff}$, $n_e (cm^{-3})$, $T_e\,(eV)$, $L_Z(\,ergs\, cm^3 s^{-1})$ and $\Delta Q_\parallel\left(\dfrac{ergs}{s\,cm^{-2}}\right)$ (6)

$$\text{In practical units}: \frac{\Delta Q_\parallel\left(\frac{GWatts}{m^2}\right)}{n_{es}\left(10^{20} m^{-3}\right)\sqrt{F_z}} \approx 2.5\times 10^5 \sqrt{T_{es}^2 \int_0^{T_{es}} L_Z(T_e) T_e^{0.5} dT_e}$$

Simple estimates by Lengyel and Lackner[64] based on the average ion impurity model[35] and later estimates based on detailed calculations by Clark et al[1] indicate that it will be difficult to radiate all of the energy for high power ITER discharges with a reasonable impurity level in coronal equilibrium (Figure 15) [1].

It is therefore important to maximize the integral $\int_0^{T'} L T^{0.5} dT$. There are two effects which can increase L(T) above the coronal equilibrium values: charge exchange recombination [48, 54], and impurity transport rapid compared to the time to approach coronal equilibrium[59, 66]. Charge exchange recombination occurs when neutral hydrogen atoms transfer their electron to an impurity ion by charge exchange producing a lower charge state of the impurity[48].

$$H^o + A^q \Rightarrow p + A^{q-1}$$
$$A^{q-1} + e^- \Rightarrow \left(A^{q-1}\right)^* + e^- \qquad (7)$$
$$\left(A^{q-1}\right)^* \Rightarrow A^{q-1} + h\upsilon$$

Although the electron is transferred into an excited state, and energy is radiated as it cascades down to the ground state, the major radiation losses are due to the excitation of ground state ions. Charge exchange recombination produces a more recombined mixture of ionic charge states which radiate more strongly than the coronal distribution of charge states. The emissivity ($P_{rad}/(n_e n_Z)$) for O is shown in Figure 16. Charge exchange recombination can increase the size of the temperature window for impurities if $n_o/n_e$ is of the order of $10^{-3}$ or greater. The major issue is quantitative one: Will the neutral density in the divertor plasma be sufficiently large to enhance the radiation losses enough to radiate the power?

The radiation loss rate can also be enhanced by the rapid recycling of impurities which transports ions from colder to hotter regions of the divertor plasma. The accurate calculation of the potential enhancement of impurity radiation due to transport effects requires solving the coupled transport equations for the dominant impurity species for all of the dominant charge states (e.g. $Be^0$, $Be^{+1}$, $Be^{+2}$, $Be^{+3}$ and $Be^{+4}$) (Eq. 5). However, the magnitude of the effect can be estimated by integrating the rate equations for different initial conditions for the case where the impurities start out as neutrals and are ionized by the plasma until they reach coronal equilibrium, and the case where they start out as fully stripped and recombine. The first case would correspond to impurities coming from the wall or cold regions of the divertor plasma to hotter regions, and the second would correspond to impurities coming from hotter regions to colder regions. The radiation is enhanced in the first case, and decreased in the second case. Since the collision rate coefficients are proportional to the electron density, the density can be factored out and equation (5) can be integrated from t=0 to t=$\tau_{recycle}$, and the results plotted as a function of $n_e \tau_{recycle}$. Figures 17 and 18 show the radiation for various values of $n_e \tau_{recycle}$ for a neutral impurity in a hot plasma and a fully stripped impurity in a hot plasma. For $n_e \tau_{recycle} \leq 10^{10}$ s cm$^{-3}$, the enhancement is appreciable for the initially neutral impurity and the effective temperature window is greatly extended. For the initially fully stripped impurity the radiation window is reduced. To show the dynamics of the situation, the history of the charge states is shown in Figures 19 and 20. Given the dramatic differences in the two cases, accurate calculations will likely require the solution of the full set of impurity rate and transport equations.

As discussed by Lackner, et al [65] and Allen, et al[66], the simple criteria developed above can be used to characterize the level of impurity recycling and charge exchange recombination required to radiate the energy. This has been done for Be, C, Ne, and Ar[67] (Figures 21 and 22 and Table 4) to determine the required neutral fraction and impurity recycling rate needed to exhaust a given power as a function of the impurity fraction, upstream density, and Zeff for Neon. The criteria can be phrased in terms of a normalized heat flux $\tilde{q}$ and distance $\xi$:

$$\tilde{q} = \frac{Q_\parallel (GWm^{-2})}{n_s(10^{20} m^{-3})} \sqrt{\frac{Z_{eff}}{f_z(\%)} \frac{\ln \Lambda}{12}} \quad and \quad \xi = n_s(10^{20} m^{-3}) \, x(100m) \sqrt{Z_{eff} f_z(\%) \frac{\ln \Lambda}{12}} \quad (8)$$

The preliminary analysis indicates that Neon is the optimum impurity for energy exhaust[67]. The increase in radiation with argon in the temperature range of interest is smaller than the decrease incurred because of the higher $Z_{eff}$ and greater core radiation. Detailed plasma modelling of the two dimensional plasma including transport effects and all of the energy loss channels are needed to draw definitive conclusions, but it appears that it should be possible to achieve radiation rate coefficients that are large enough to exhaust ~ 200 MW from the ITER divertor if $n_o/n_e > 10^{-2}$ and/or $n_e \tau_{recycle} < 3 \times 10^{10}$ s cm$^{-3}$.

Table 4  Comparison of Be, C, Ne, and Ar radiation efficiencies for DIII-D and ITER.[67]

| Element | Be | C | Ne | Ar |
| --- | --- | --- | --- | --- |
| $0.33 \times$ fatal $f_z(\%)$ | 4.7 | 2.23 | 0.8 | 0.18 |
| $\sqrt{(f_z(\%)/Z_{eff})}$ | 1.73 | 1.16 | 0.68 | 0.34 |
| $Q_{\parallel DIII-D}/\sqrt{(f_z(\%)/Z_{eff})}$ | 0.27 | 0.41 | 0.69 | 1.38 |
| $n_o/n_e$ | $5 \times 10^{-2}$ | $10^{-2}$ | $10^{-2}$ | $5 \times 10^{-2}$ |
| $n_e \tau_{recy}$ (s cm$^{-3}$) | $10^{10}$ | $10^{10}$ | $3 \times 10^{10}$ | $5 \times 10^9$ |
| $Q_{\parallel ITER}/\sqrt{(f_z(\%)/Z_{eff})}$ | 0.87 | 1.3 | 2.2 | 4.4 |
| $n_o/n_e$ | $7 \times 10^{-3}$ | $8 \times 10^{-3}$ | $10^{-3}$ | $4 \times 10^{-2}$ |
| $n_e \tau_{recy}$ (s cm$^{-3}$) | $10^{10}$ | $10^{10}$ | $4 \times 10^{10}$ | $6 \times 10^9$ |

Data for hydrocarbon reactions is of interest because of the use of graphite plasma facing components in many experiments [68, 69]. Chemical sputtering is an important erosion mechanism[70]. Atomic hydrogen is quite reactive and will produce some hydrocarbon flux at the plasma edge from graphite plasma facing components. A dataset was put together in 1987 to model carbon influxes into experiments[71]. This database has been extended and reviewed[72] so that quantitative modelling can be done for the whole reaction chain starting with methane, ethane, ethene, acetylene, etc., reacting with electrons to yield in the plasma H, $H^+$, C and $C^+$.

## 4. Databases and Datacenters

During the last twenty or so years, an active group of data centers dedicated has been collecting and analyzing atomic physics data for fusion (Table 5)[31, 73, 74, 75]. Because fusion researchers do not usually read the world wide atomic physics literature, much of this activity has concentrated on compiling and classifying bibliographies of the atomic physics literature so that the relevant articles and data can be accessed by fusion scientists. Most of these data centers have also collected, assessed, evaluated, and processed data for atomic collisions, wavelengths and transition strengths published in reports such as the ORNL Redbooks (e.g. [76, 77, 78, 79]). These groups have served as useful sources of data for the fusion community. Because most of the data centers are part of strong atomic physics groups, they also help to keep the atomic physics community aware of and working on some of the key atomic physics questions important for fusion. As described in previous sections, there has been considerable progress in the ability to accurately calculate ionization, recombination, and excitation rate coefficients for most of the ions of interest to the fusion community, especially those important at the plasma edge.

Up to this time, the data has often not been easily accessible to the general fusion scientist. The situation should soon improve dramatically because many of the groups are putting their databases with software on local UNIX workstations which will be accessible through the INTERNET by the general fusion community. This access will make it possible for both the specialized and general fusion scientist to get data and perform bibliographic searches from a terminal connected to INTERNET. Most of these groups have standardized on the ALADDIN database format for the data adopted by the IAEA [80]. These groups have made a determined effort to evolve their data collections to reflect the shifts in the priorities of the fusion community. The emphasis of many of them is now beginning to shift from the plasma core toward the plasma edge. Low temperature impurity rate coefficients, elastic scattering and molecular effects are beginning to receive a lot of attention[81]. An effort is also beginning to collect and analyze plasma surface and materials data[82, 83].

Table 5. Representative List of Atomic and Molecular Data Centers for Fusion[73]

| Location | Contact Person | Type of data | Access and database format |
|---|---|---|---|
| International Atomic Energy Agency | R. Janev J. Botero: rnd@iaea1.or.at | Bibliography Cross sections Reaction Rate coefficients | ALADDIN, UNIX workstation on INTERNET aladdin@ripcrs01.iaea.or.at |
| National Institute of Standards and Technology (NIST) Washington, DC USA | W. L. Wiese | Bibliography Spectroscopic data (wavelengths, transition probabilities, line shapes, etc.) | UNIX ORACLE system on INTERNET |
| A&M Data Center at the Kurchatov Institute, Moscow, RF | V. Abramov krash@kiae.su | atomic collision data (Be, Ga, etc.) physical sputtering | |
| Max Planck Institute for Plasma Physics, Garching | W. Eckstein: wge@ipp-garching.mpg.de | neutral reflection and sputtering data | by mail |
| Joint Institute for Laboratory Astrophysics (JILA), U. of Colorado, Boulder, Colorado | J. Broad | atomic collision data, bibliography, laser orientation-wavelengths, transition strengths | DEC 5000/200 UNIX with INGRES, jtb@jiladc.colorado.edu |
| Centre de Donnees Graphyor, ORSAY, France | J. Delcroix | collision data, bibliography, energy levels,.. | being migrated to a UNIX workstation |
| National Institute for Fusion Science, Nagoya, Japan | H. Tawara T. Kato: takako @nifs.ac.jp | bibliography, atomic collision data, | ALADDIN, UNIX workstation on INTERNET: msp.nifs.ac.jp |
| Oak Ridge National Laboratory Controlled Fusion Atomic Data Center | D. Schultz schultz@orph28. phy.ornl.gov | bibliography, atomic collision data | ALADDIN, migration to UNIX workstation in progress |
| Nuclear Data Center, JAERI, Tokai, Japan | T. Shirai: j3323@jpnjaeri | bibliography, atomic collision data | |
| Atomic and Molecular Data Unit, Queen's University of Belfast, Northern Ireland | S. Saadat F. Smith | bibliography, atomic collision data | |
| Joint European Torus, Abingdon, UK | H. P. Summers | atomic collision data, primarily for diagnostics | ADAS system on IBM 3090, migration to UNIX workstation in progress |
| Laboratoire de Physique des Gas et des Plasmas, Universite de Paris-Sud Batiment 212 F-91405 Orsay Cedex, France | K. Katsonis | Cross sections | |

## 5. Summary

As divertor research evolves to address the problems of exhausting very large power level for next step experiments, atomic processes have been emphasized as the main method for removing the power from the plasma. Because of this, there is renewed interest in hydrogen and impurity atomic physics at low temperatures (0.5—300 eV) and moderate to high densities ($10^{19}$—$10^{22}$ m$^{-3}$). Collisional radiative effects are important for determining the hydrogen ionization and recombination balance and the hydrogen radiation. Line transfer effects may also be important, especially for the ionization/recombination balance. Molecules, including vibrationally excited molecules, will also be important. Negative ions may play a significant role in the particle balance by offering a potential recombination mechanism. Elastic scattering of atoms, ions, and molecules also can play an important role.

Impurity radiation offers the best promise for a way to spread out the plasma heating energy over a large area and reduce the peak heat loads. The emphasis in impurity atomic physics has shifted from the high temperature region in the plasma core to the low temperature conditions appropriate for divertor operation ($\leq \sim 300$ eV). There has been considerable progress in our knowledge of the basic ionization, recombination, and excitation rate coefficients. Impurity radiation will probably be too feeble unless it is enhanced by such processes as charge exchange recombination and impurity recycling.

All of these processes will need to be included in the comprehensive computational models being developed for analyzing the experiments and extrapolating to the next generation of tokamak experiments. The atomic data being developed is being assembled at atomic data centers around the world for use by the fusion community. This data should shortly be routinely available to the entire fusion community via INTERNET.

## Acknowledgments

I am grateful for encouragement and discussions with Drs. R. Janev, S. Allen, J. Botero, R. Clark, G. Dunn, D. Hill, R. Hulse, G. Janeschitz, K. Lackner, F. Perkins, M. Petravic, R.

Phaneuf, M. Pindzola, N. Putvinskaya, P. H. Rebut, D. Reiter, M. Rosenbluth, D. Schultz, D. Stotler, H. Summers, H. Tawara, A. Wan and J. Weisheit.

Figures:

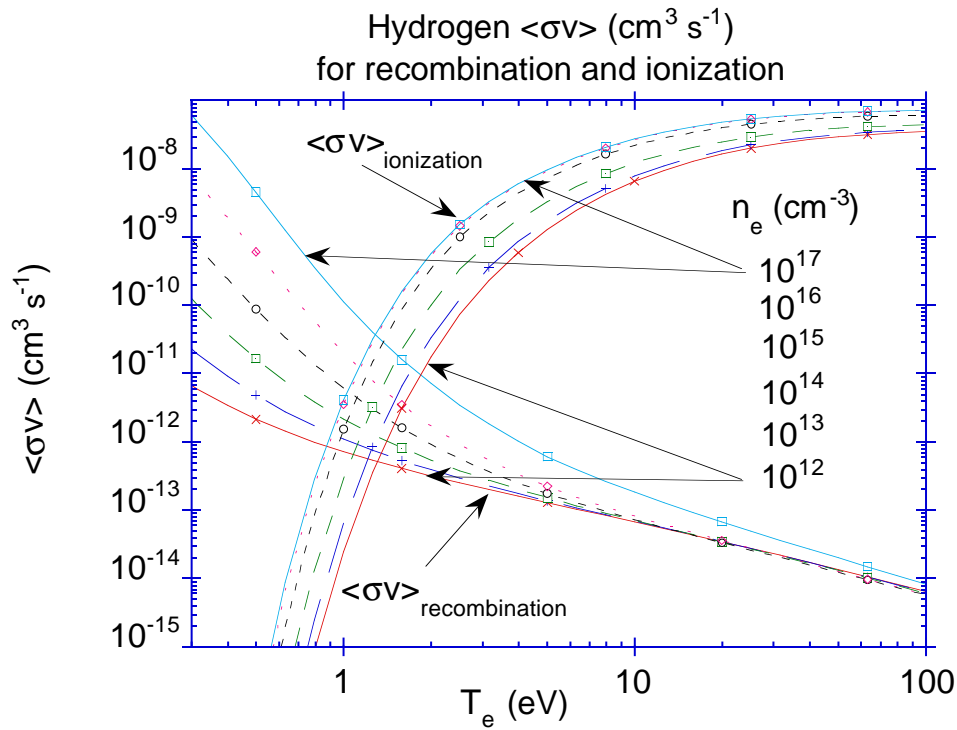

Figure 1 Hydrogen Ionization and Recombination

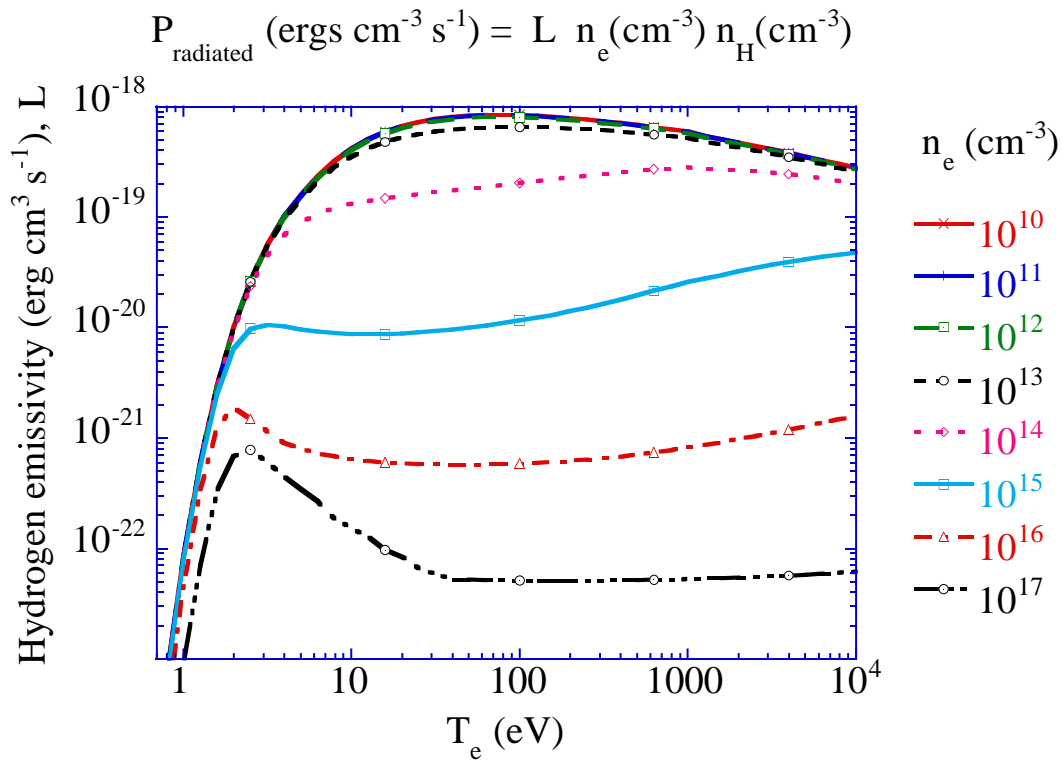

Figure 2 Hydrogen Radiation Emission

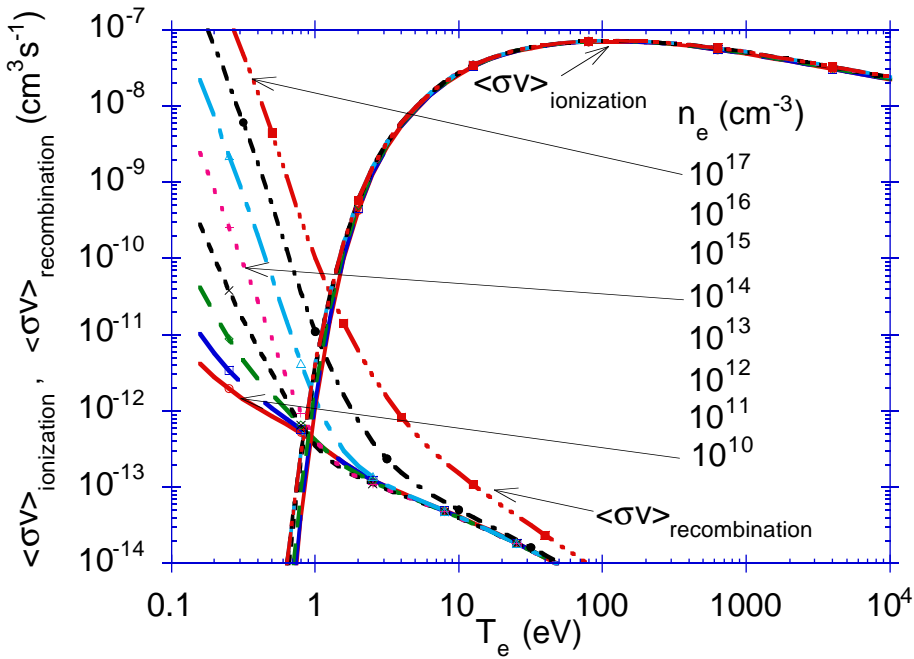

Figure 3 Case B[13]: H ionization / recombination

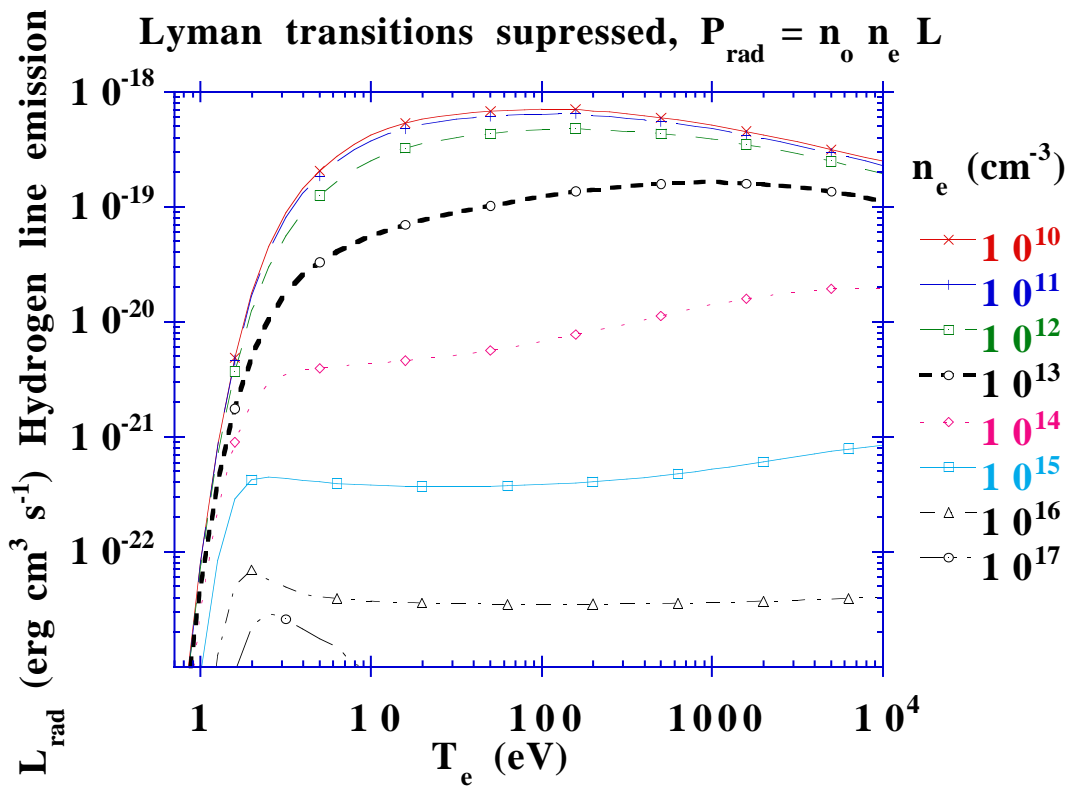

Figure 4. Case B[3]: H radiation[13]

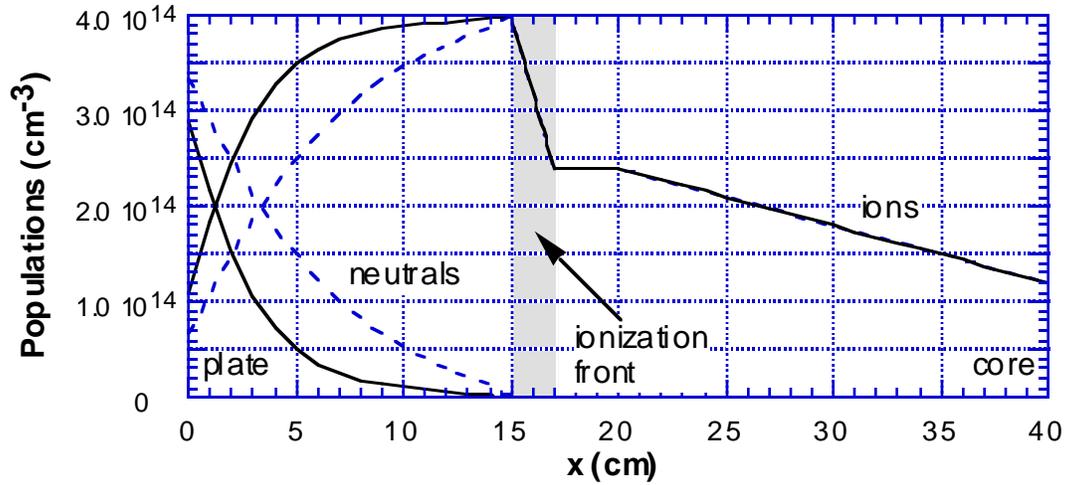

Figure 5 Spatial profiles of the neutral and ion densities for case (2), in solid lines, and case (1), in dashed lines. The inclusion of a strong, optically thick Lyman-α line of case (2) increases the ion population due to the combined processes of first excitation of the ground state electrons and subsequent collisional ionization[4].

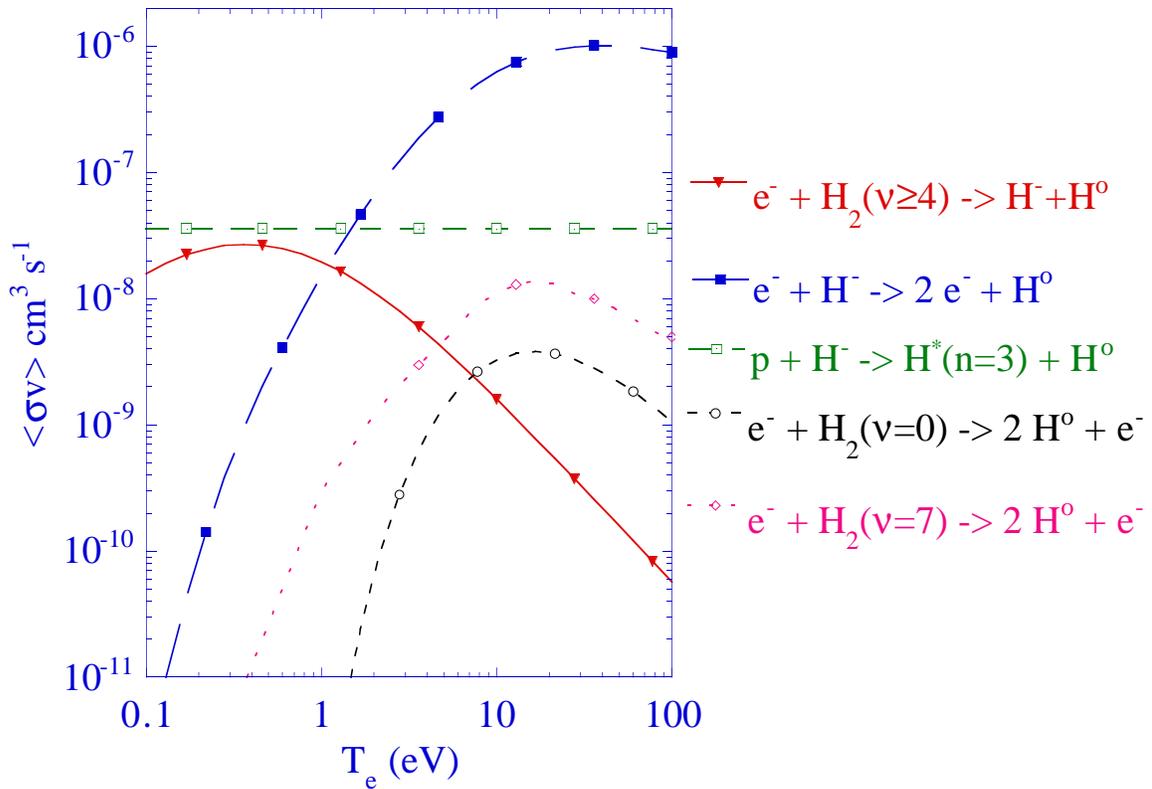

Figure 6 Rate coefficients for recombination and detachment processes for H$^-$.[16]

Shortened PSI Review Paper 6          23

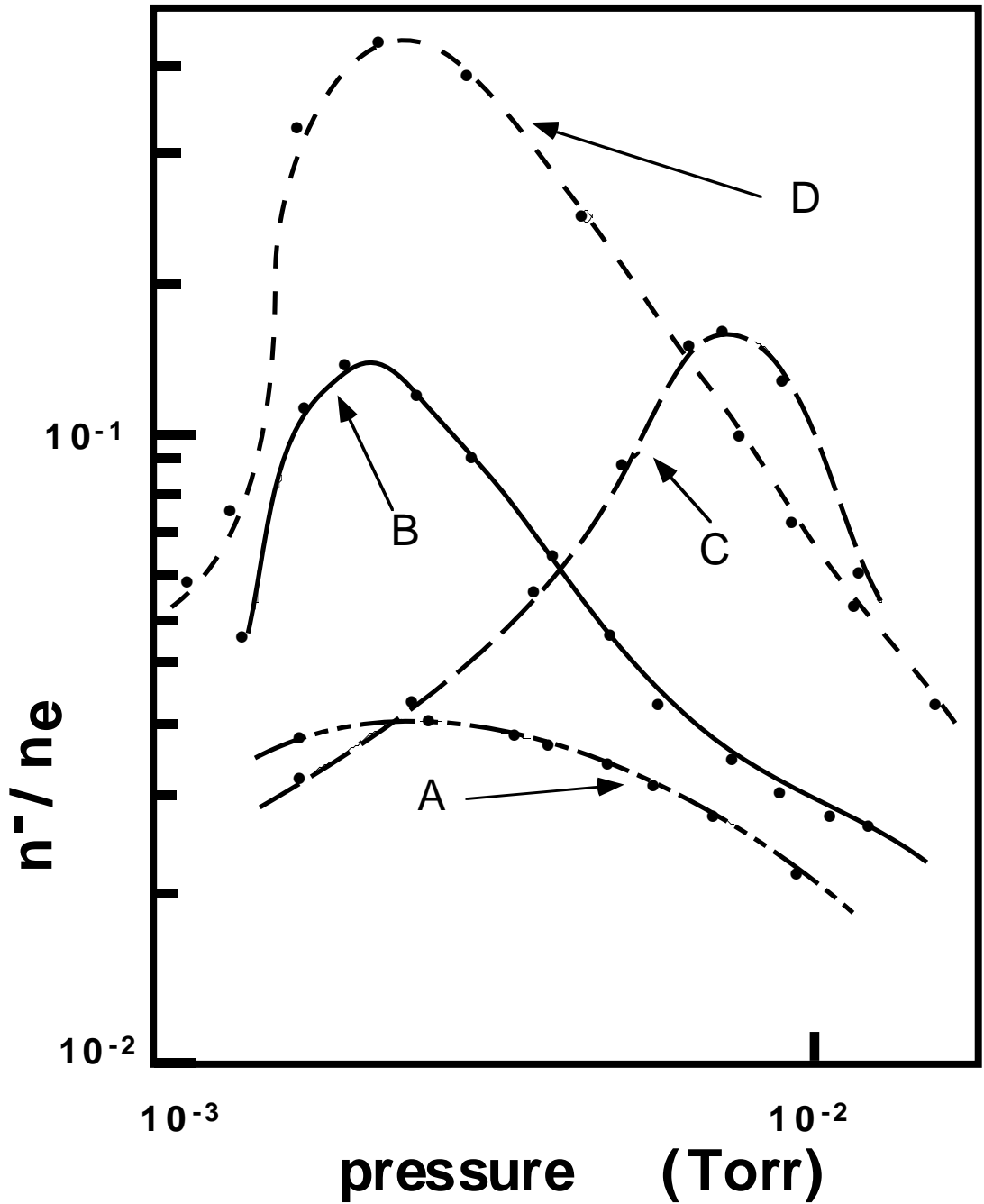

Figure 7. Ratio of H- density to $n_e$ as a function of pressure for four different cusp configurations A, B, C, and D[23].

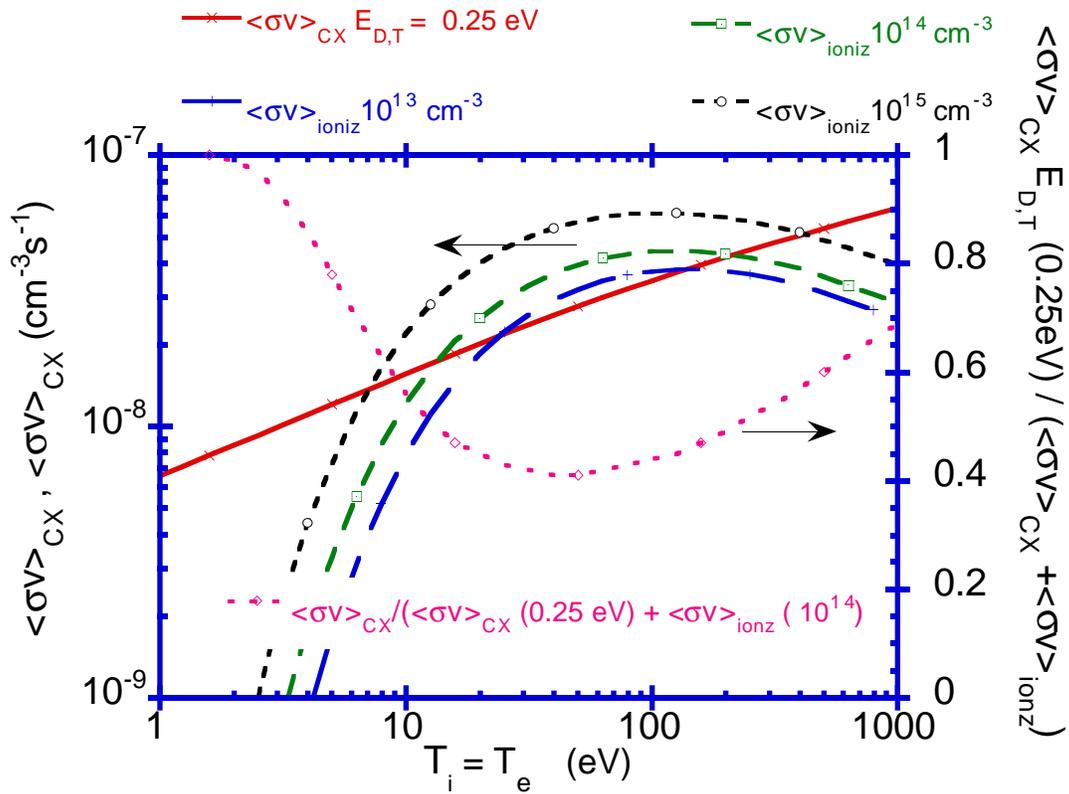

Figure 8 Comparison of the collisional radiative ionization rate and the charge exchange rate[2]

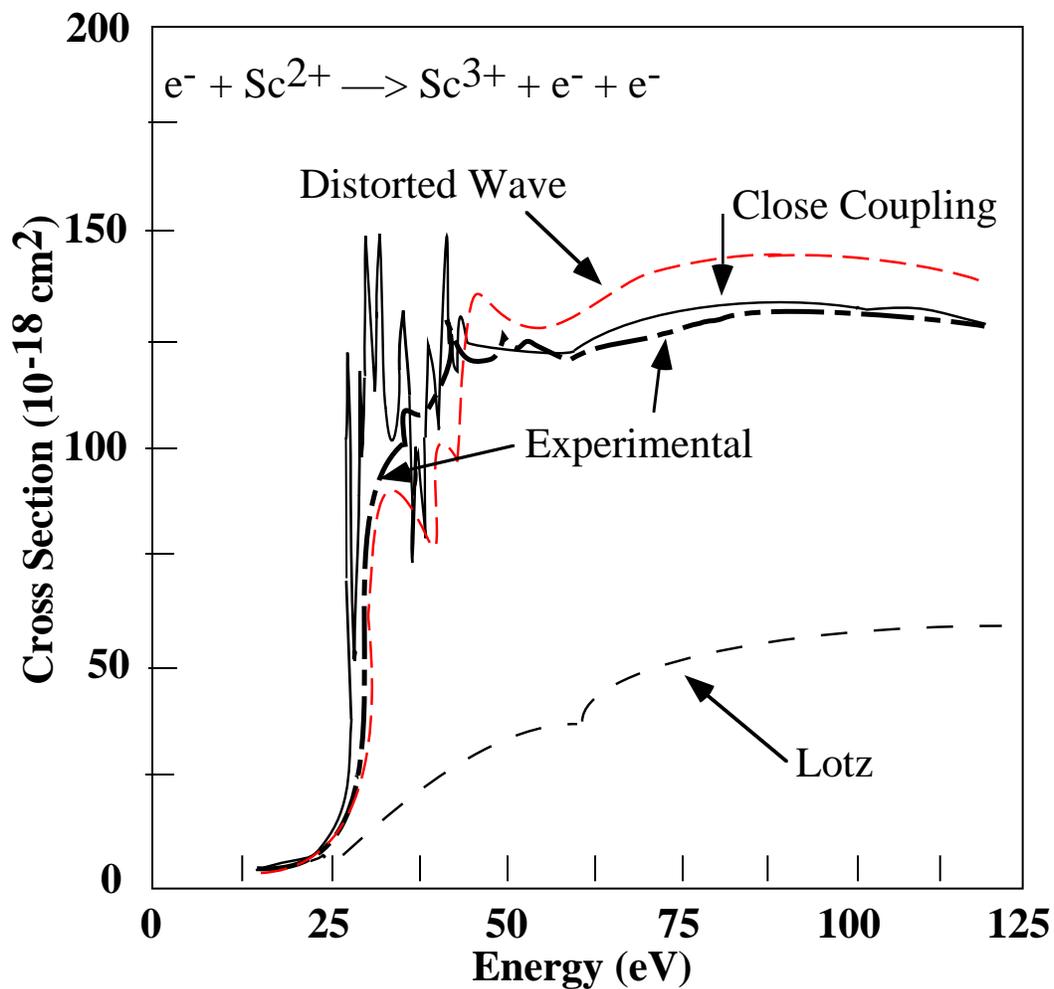

Figure 9. Comparison of the measured and calculated electron impact ionization cross section of $Sc^{2+}$ [46]. The cross section includes a strong contribution from indirect processes, including inner shell excitations. The top dotted curve is a Distorted Wave calculation, the bottom dashed curve is the Lotz semi-empirical fit for direct ionization, and the middle solid curve is from a close coupling calculation.

Shortened PSI Review Paper 6    26

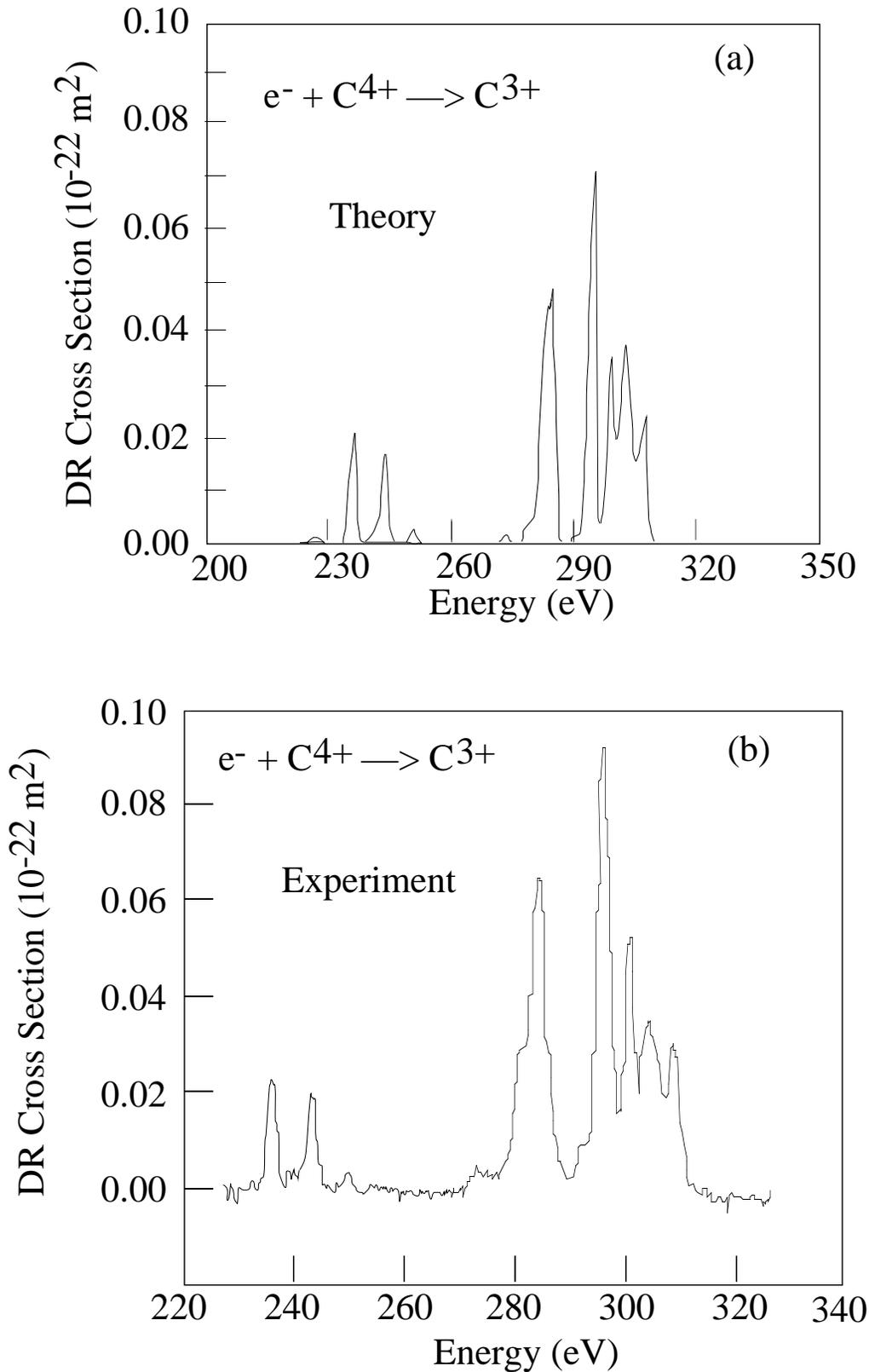

Figure 10. A comparison of the calculated (a)[84] and measured (b)[85] dielectronic recombination of $C^{4+}$.

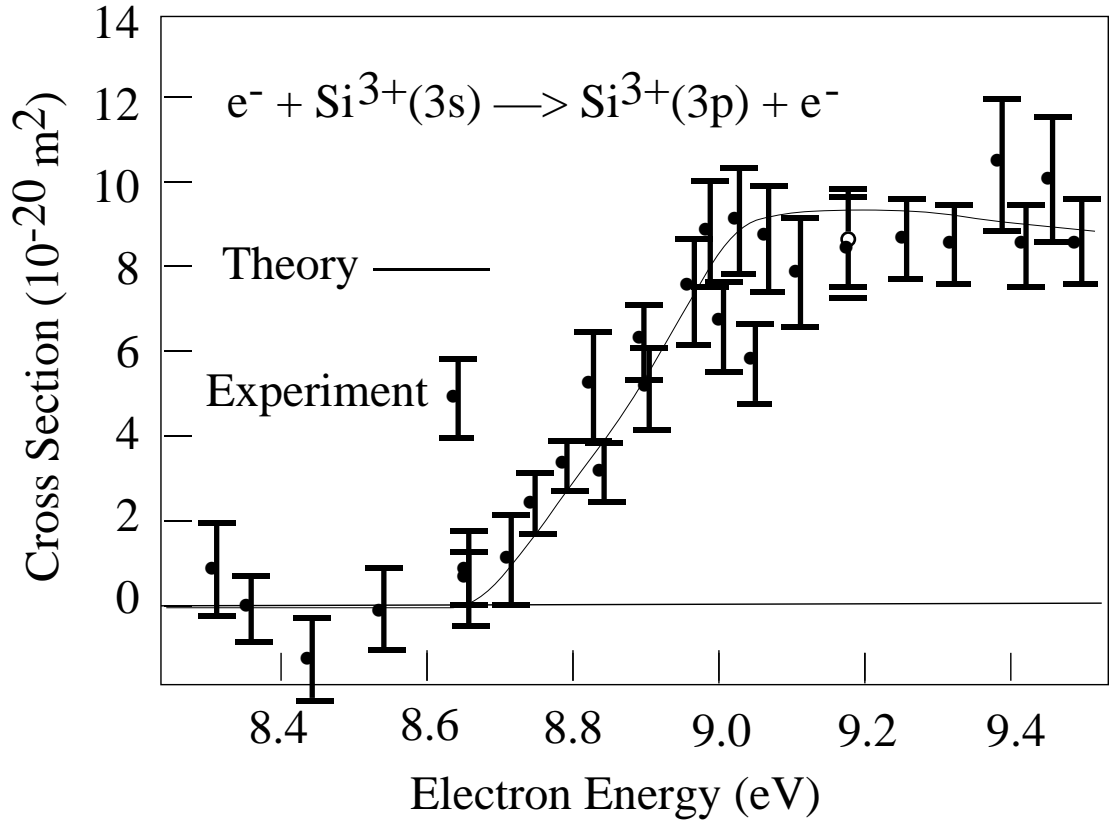

Figure 11. A comparison of the measured[56] and calculated [58] electron impact excitation cross section for $e^- + Si^{3+}(3s) \longrightarrow Si^{3+}(3p) + e^-$.

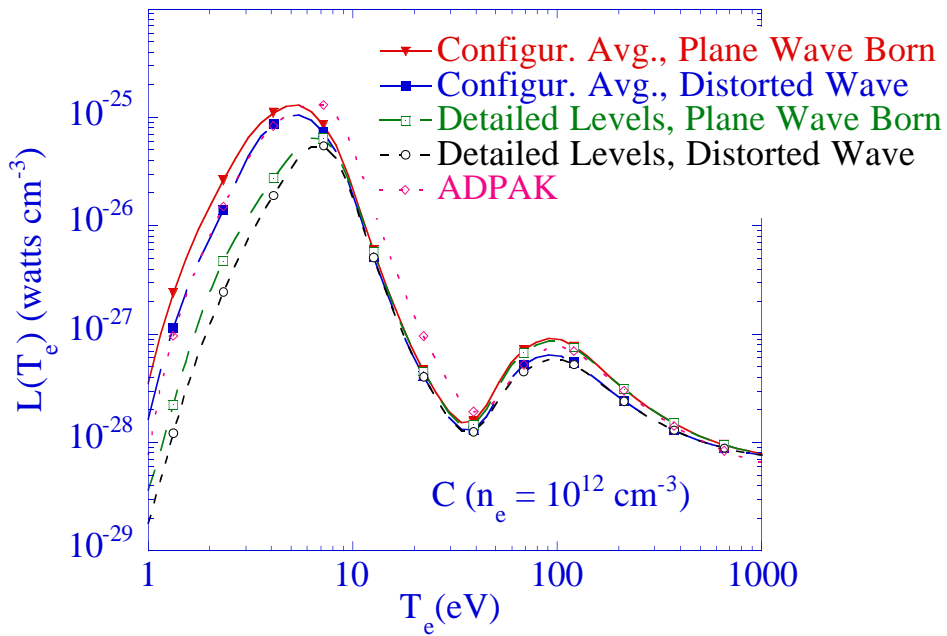

Figure 12. Comparison of Carbon Emission Rate coefficients for four models with $n_e = 10^{12}$ cm$^{-3}$.

Shortened PSI Review Paper 6       28

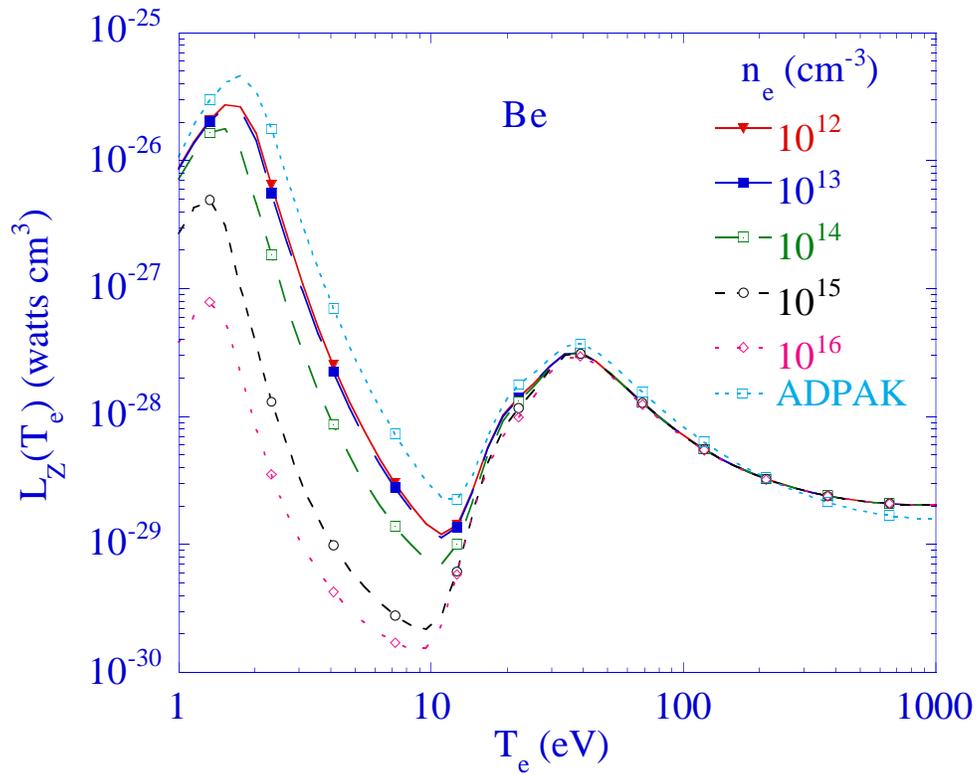

Figure 13. Collisional-Radiative emission rate coefficients for Be for $n_e = 10^{12}$—$10^{16}$ cm$^{-3}$[1]

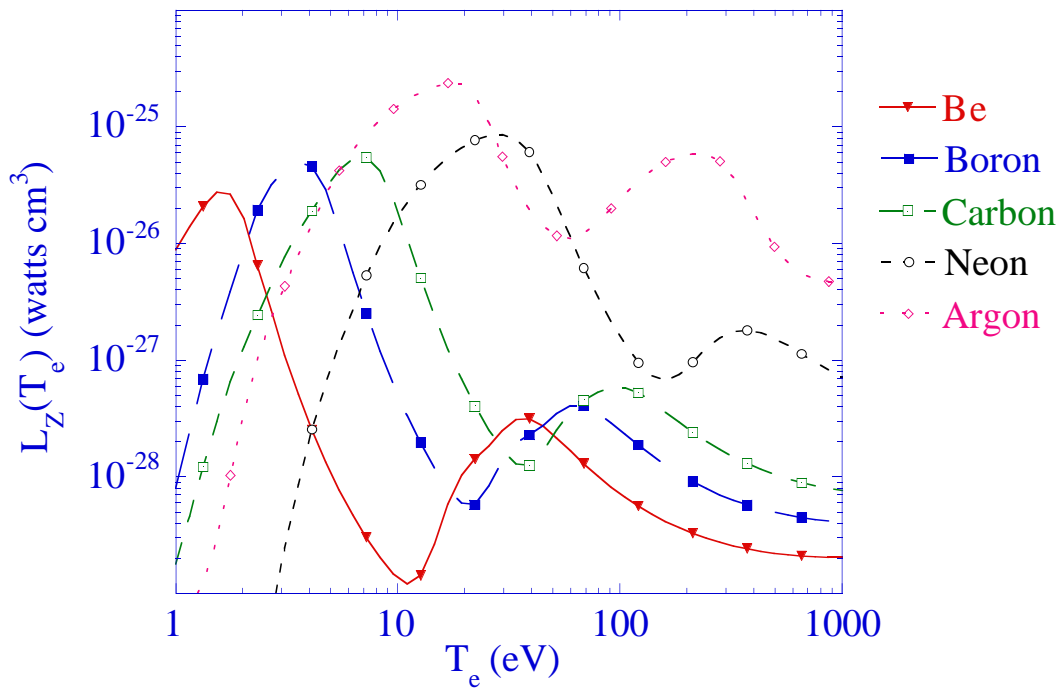

Figure 14 Collisional radiative power loss emission rate coefficients for Be, B, Carbon, Neon, and Argon for a density of $10^{12}$ cm$^{-3}$ ($P_{loss} = n_e n_z L_z(T_e)$).[1]

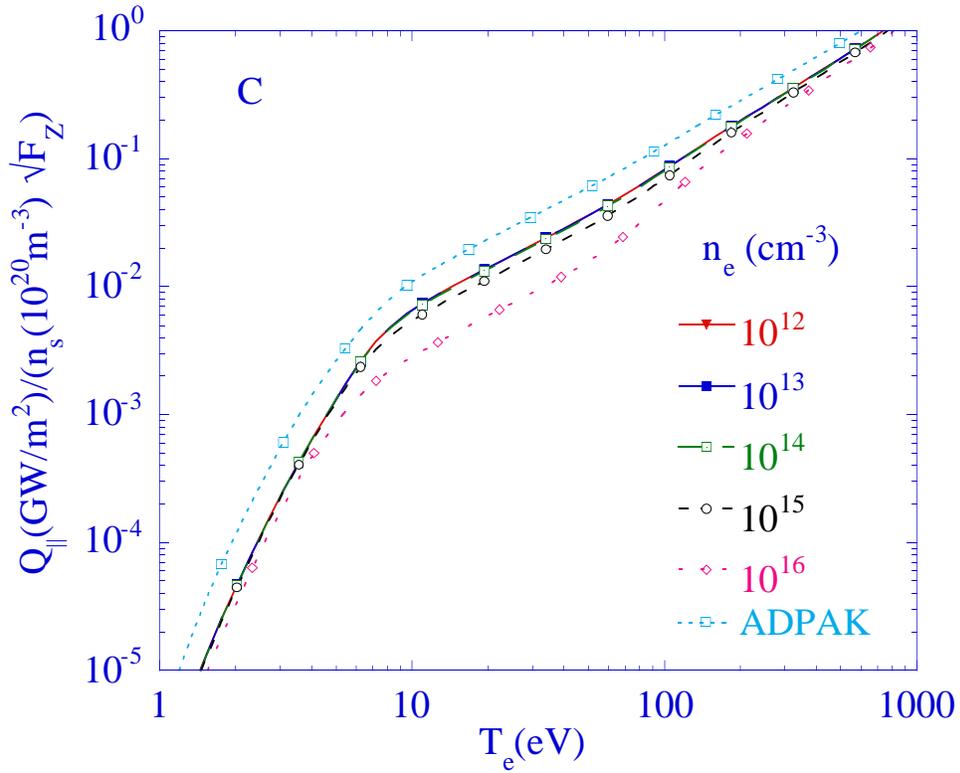

Figure 15 Carbon Cooling Rate efficiency[1]

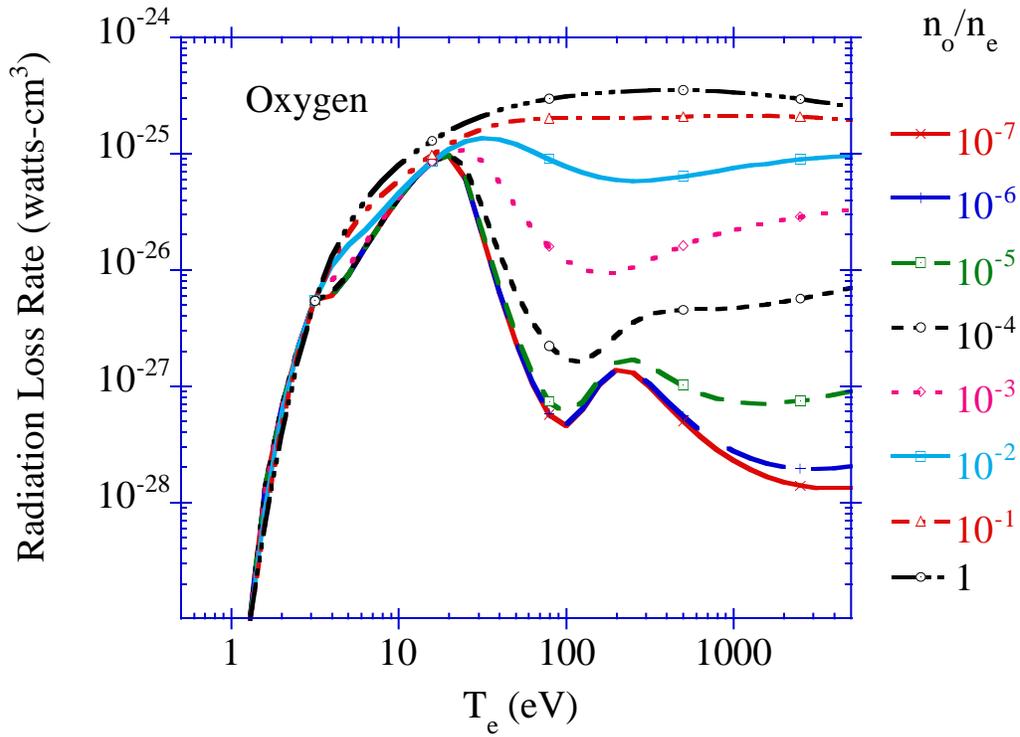

Figure 16 Oxygen radiation rate coefficients enhanced by charge exchange recombination

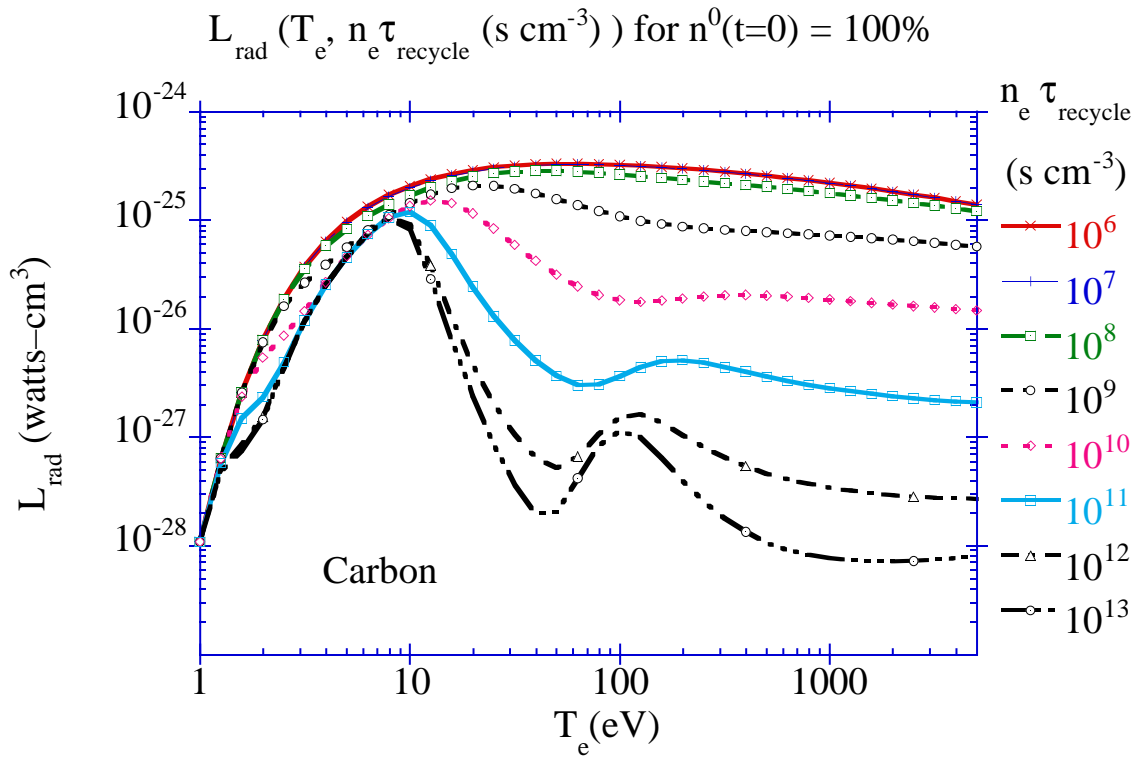

Figure 17 Emissivity of Ionizing $C^o$

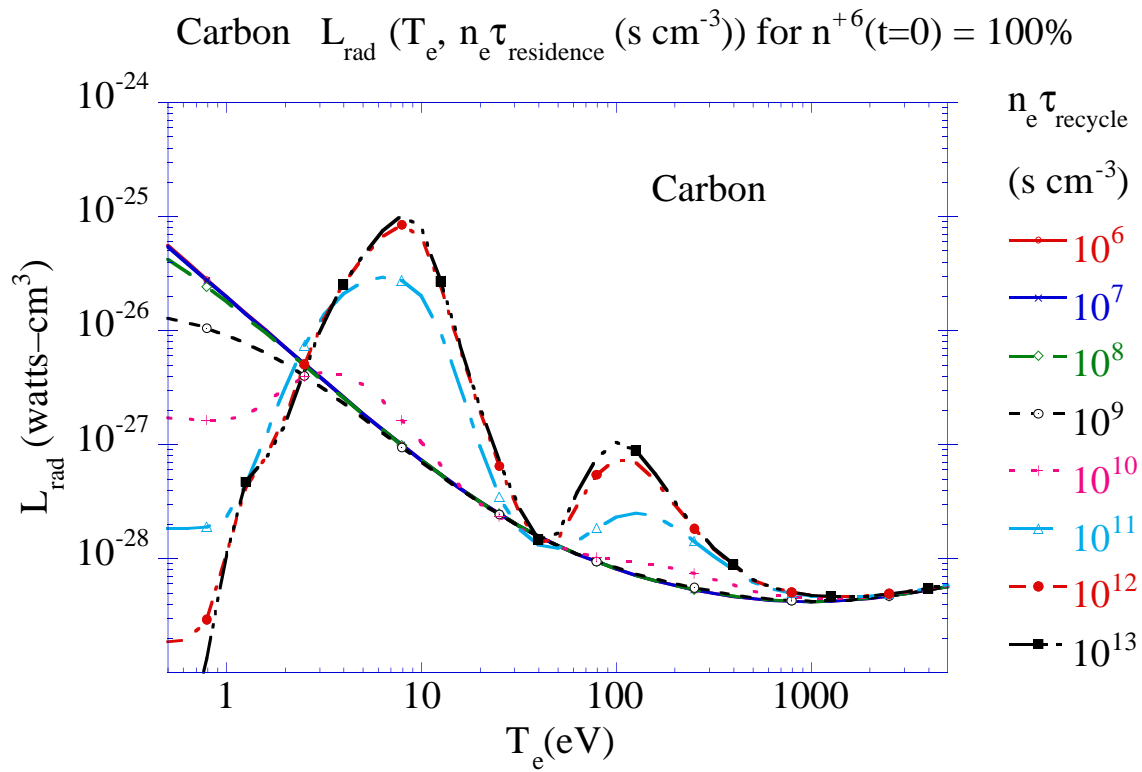

Figure 18 Emissivity for recombining $C^{+6}$

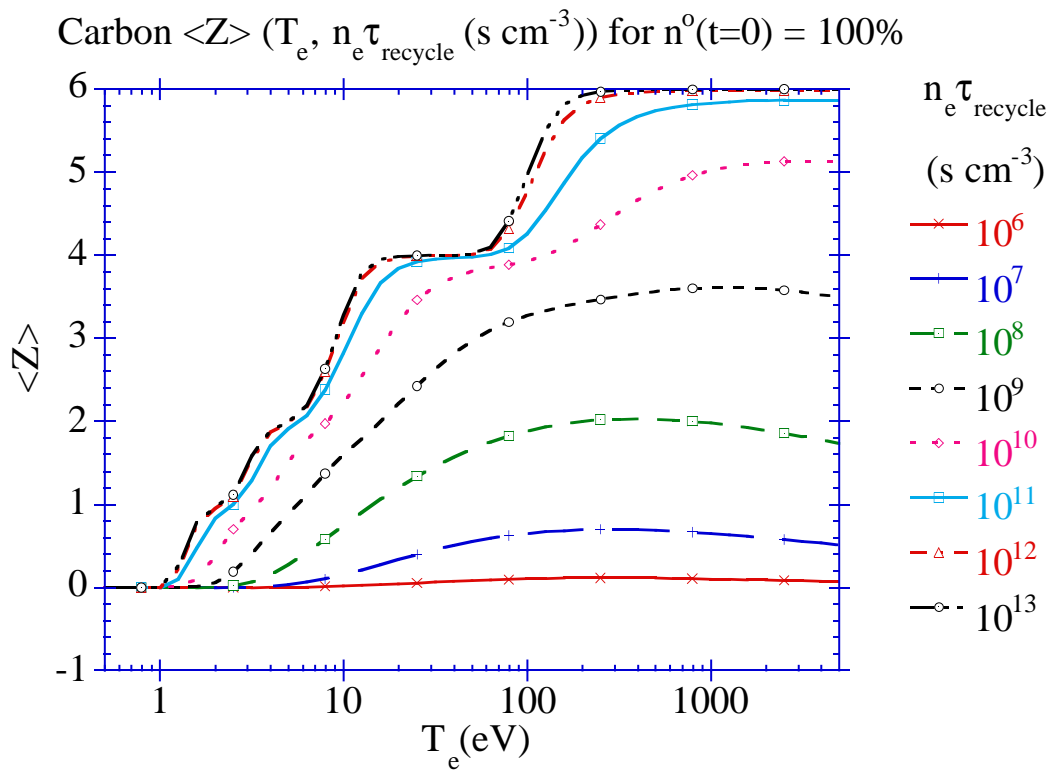

Figure 19 <Z> for ionization of $C^o$ with $n^{+0}(t=0) = 100\%$.

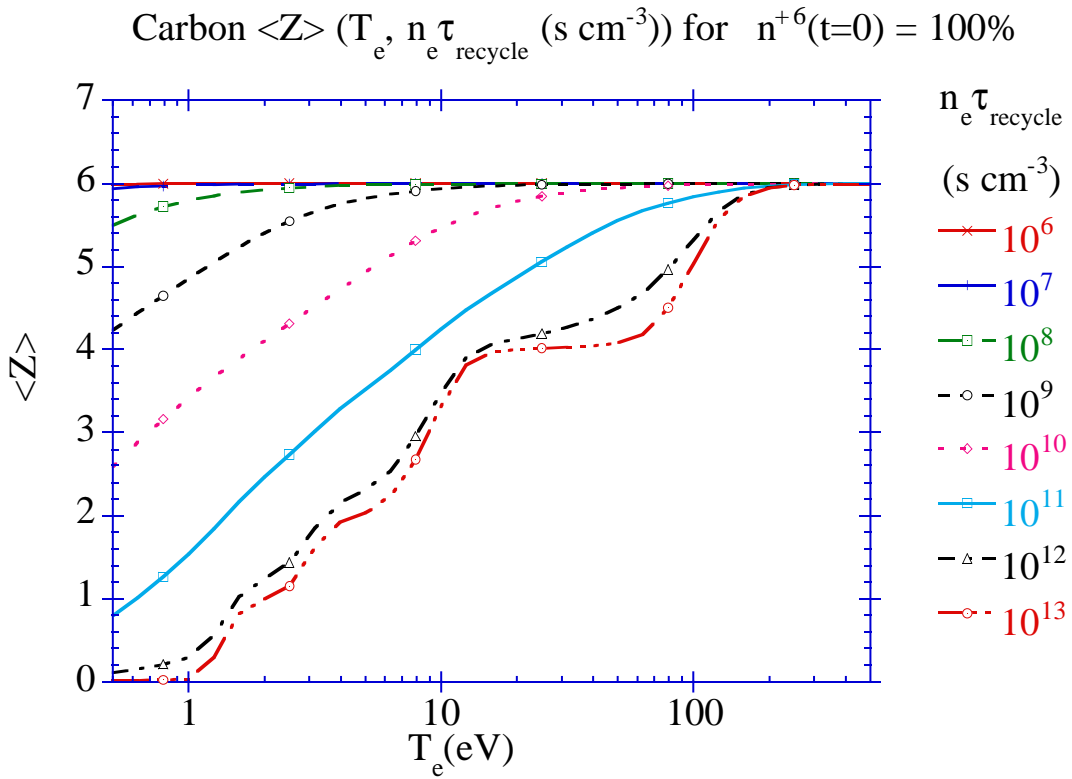

Figure 20 <Z> for recombining $C^{+6}$ with $n^{+6}$ (t=0) = 100%.

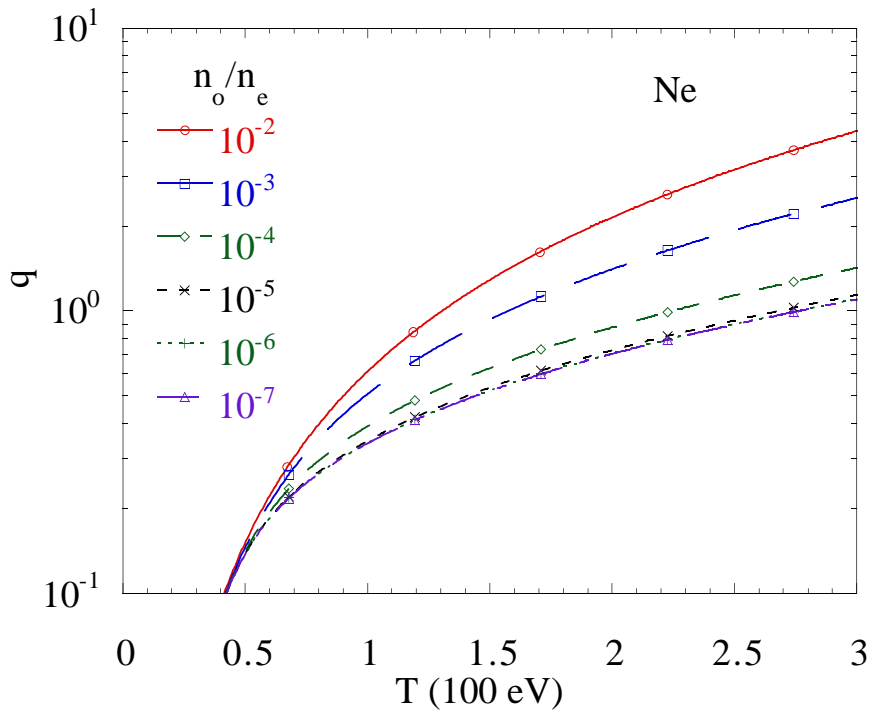

Figure 21 Required neutral fraction and impurity recycling rate to radiate a given power with a given impurity concentration and upstream density for Ne[67].

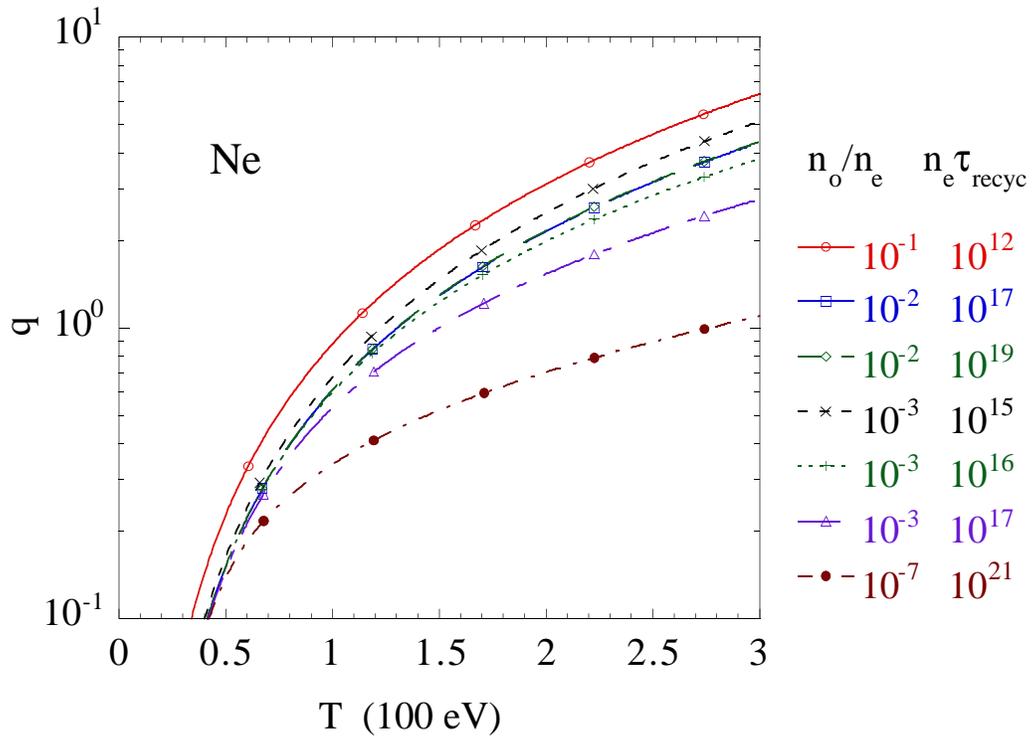

Figure 22 Required neutral fraction and impurity recycling rate to radiate a given power with a given impurity concentration and upstream density for Ne[67].